\documentclass[conference]{IEEEtran}
\usepackage{cite}
\usepackage{amsmath,amssymb,amsfonts}
\usepackage{graphicx}
\usepackage{textcomp}
\usepackage{xcolor}
\usepackage{booktabs,array,tabularx}
\usepackage{tikz}
\usetikzlibrary{arrows.meta,positioning,calc}
\usepackage[hidelinks]{hyperref}

\hypersetup{pdftitle={Point-Supervised Visual Assessment of GNSS Measurement Quality with Automatic Multi-Reference Labeling},pdfkeywords={GNSS, DGNSS, fisheye imaging, sparse point supervision}}

\definecolor{draftred}{RGB}{150,38,38}
\definecolor{flowblue}{RGB}{38,77,112}
\definecolor{flowgreen}{RGB}{37,99,80}
\definecolor{flowgold}{RGB}{135,94,35}

\newcolumntype{L}[1]{>{\raggedright\arraybackslash}p{#1}}
\newcolumntype{Z}{>{\raggedright\arraybackslash}X}

\author{%
\IEEEauthorblockN{%
Weiwei Zhai\textsuperscript{1}, 
Penggao Yan\textsuperscript{1,*},
Liang Wang\textsuperscript{2}, and
Li-Ta Hsu\textsuperscript{1}, \textit{Senior Member, IEEE}}
\IEEEauthorblockA{%
\textsuperscript{1}Department of Aeronautical and Aviation Engineering,
The Hong Kong Polytechnic University, Hong Kong SAR, China\\
\textsuperscript{2}Aerospace Information Research Institute, Chinese Academy of Sciences, Beijing 100094, China\\[2pt]
\textsuperscript{*}Corresponding author: Penggao Yan\\
\texttt{peng-gao.yan@connect.polyu.hk}
}}

\usepackage{microtype}               
\begin{document}

\title{Point-Supervised Visual Learning for GNSS Measurement Quality Assessment}
\maketitle

\begin{abstract}
Multipath and non-line-of-sight (NLOS) propagation degrade global navigation satellite system (GNSS) positioning accuracy in urban environments.
Although fisheye sky segmentation can identify the obstructions in the direction of satellite signal propagation, binary visibility is still insufficient to fully characterize the observation quality. 
Therefore, this study proposes a point-supervised visual framework that utilizes the supervisory information obtained from GNSS observations and learns to predict GNSS observation faults from fisheye images.
During offline labeling, a vision-assisted differential GNSS procedure selects multiple reference satellites based on sky visibility and elevation constraints. 
The double difference pseudorange residuals are used to automatically generate fault labels and sample importance weights at satellite projection locations.
During training, the residual weighted loss is calculated only at satellite projection points with valid labels, allowing the network to learn the relationship between the visual environment and GNSS observation faults from sparse point labels without manual pixelwise fault annotations. The trained network generates a dense fault probability map from each fisheye image. The predicted probabilities are then sampled at satellite projection locations and used to adjust observation weights in weighted least squares positioning, reducing the influence of potentially faulty observations on the positioning solution.
The framework is evaluated using U-Net, DeepLabv3+, DINOv3, and SAM3 on urban vehicle data collected on a different day from the training data. All four models reduce horizontal root mean square error (RMSE) compared with SAM binary sky segmentation, with reductions ranging from 29.51\% to 36.07\%. U-Net achieves the lowest horizontal RMSE of 0.78 m. The framework also provides positioning solutions during some intervals when the segmentation baseline fails.
These results demonstrate the generalization ability of the proposed framework and show that point supervision derived from GNSS observations enables fisheye images to provide observation quality information beyond binary sky visibility, supporting GNSS fault detection.

\end{abstract}


\section{Introduction}

Global navigation satellite systems (GNSS) provide positioning, navigation and timing (PNT) services for vehicle navigation and autonomous driving applications \cite{10040628}.
However, reliable GNSS positioning remains challenging in dense urban environments\cite{adjrad2018intelligent}.
Buildings, vegetation, and overpasses can obstruct direct satellite signals or reflect them, leading to multipath interference and non-line-of-sight (NLOS) reception \cite{10516283}.
These propagation effects can introduce pseudorange measurement errors, thereby degrading GNSS positioning accuracy.
Therefore, accurately identifying GNSS observations affected by multipath and NLOS reception is important for improving positioning accuracy and supporting integrity monitoring\cite{11554454}.

To mitigate the effects of multipath interference and NLOS reception on positioning accuracy, GNSS observation quality control has been widely studied.
Traditional methods commonly use satellite elevation angles\cite{Li2023} and the carrier-to-noise density ratio ($C/N_0$)\cite{Strode2016} to assess observation quality and assign weights.
However, in urban environments, signals from satellites at high elevation angles may still be obstructed by tall buildings. Therefore, a higher elevation angle does not necessarily imply better observation quality.
Therefore, these indicators have limited ability to characterize the observation quality in urban environments \cite{Li2023MLGNSS}.
Building on these approaches, receiver autonomous integrity monitoring (RAIM) detects faults by checking the consistency of redundant observations using residuals\cite{Hewitson2006}.
Sequential exclusion and comparisons of solutions obtained from different observation subsets can further identify and exclude suspected faulty observations\cite{joerger2014solution}.
However, when multiple satellite observations are faulty, the receiver clock bias may be incorrectly estimated, causing fault-free observations to be mistakenly identified as faulty\cite{zhai2020integrated}.
Advanced receiver autonomous integrity monitoring (ARAIM) uses subset monitoring to account for hypotheses involving multiple simultaneous faults\cite{7472967}.
Nevertheless, the number of subsets to be monitored grows combinatorially with the number of satellites and the maximum number of simultaneous faults considered, increasing the computational burden for positioning in real time\cite{Meng2019}.

In recent years, machine learning methods have also been applied to identify GNSS observations affected by multipath and NLOS \cite{10516283}.
A Transformer-enhanced long short-term memory (LSTM) network integrates temporal features of GNSS observations with spatial contextual information from multiple satellites for NLOS detection and pseudorange error prediction\cite{10422672}.
Additionally, the method based on LightGBM integrates multiple GNSS signal features to identify and eliminate NLOS observations \cite{Zhang2024}.
To improve generalization across environments, a spatiotemporal cross-attention model combines airspace attention, temporal feature extraction using LSTM, and sparse regularization to improve NLOS classification accuracy across different locations\cite{10586781}.

However, these methods mainly rely on GNSS observations, and their fault detection performance is still limited by the ability of the used features to distinguish between healthy observations and faulty observations.
When these features do not provide sufficient information for fault detection, further algorithmic optimization may yield only limited improvements in detection performance.
To provide complementary environmental information, visual information has been introduced into GNSS observation fault detection.
Fisheye cameras are low-cost devices that capture the visible sky and surrounding obstructions.
Image segmentation generates a sky mask by distinguishing sky regions from obstacles.
Satellite directions are then projected onto the image to identify NLOS signals. 
The method using the Swin Transformer was proposed to segment fisheye images into sky and non-sky regions for NLOS detection\cite{10422540}. 
In addition, a method combining DeepLabv3+ sky segmentation with building edge geometry was proposed to identify NLOS observations. Based on the relative positions of satellite directions and building edges, an observation weighting strategy was developed to mitigate the impact of NLOS propagation on positioning accuracy\cite{Liu2025}.

Visual visibility does not uniquely determine GNSS observation quality. 
Even when a direct signal is available, an LOS observation may still be affected by multipath if reflected signals are received simultaneously\cite{Groves2013}.
Additionally, the NLOS ranging error varies with the geometric relationship between the satellite and the reflecting surface, and its magnitude cannot be solely determined by the NLOS category \cite{ Hsu2018}.
Therefore, although binary sky segmentation provides information about signal obstruction, it cannot reliably identify faulty GNSS observations. 
Then, can we utilize the supervisory information generated by GNSS observations themselves to enable the \textbf{visual model} to acquire an observation fault detection capability \textbf{beyond the binary sky visibility?}

To investigate this question, this study develops a visual framework for GNSS observation fault detection using point supervision.
During training, automatically generated differential GNSS (DGNSS) residuals obtained using multiple references are used to assign fault labels at satellite projection locations in the images.
These labels guide the model to learn the relationship between visual features and GNSS observation faults.
The objective is to predict GNSS observation fault probabilities from fisheye images without manual pixelwise fault annotations.

The main contributions of this study are summarized as follows:
\begin{enumerate}
    \item The point supervision framework is employed to generate fault probability maps from fisheye images, in order to reflect the influence of the surrounding environment on the quality of GNSS observations. Compared with binary sky segmentation, these maps provide more information about observation quality.
    \item We generate point labels and training weights automatically from DGNSS observations using multiple reference satellites selected according to sky visibility and elevation. After removing the geometric range differences, the residuals define observation faults and their importance in training. The loss is calculated only at labeled satellite positions, removing the need for manual pixelwise fault annotations.
    \item Vehicle experiments with U-Net, DeepLabv3+, DINOv3, and SAM3 show that using the predicted fault probabilities to weight GNSS observations reduces horizontal positioning errors compared with SAM sky segmentation. The framework also provides positioning solutions during some periods when the segmentation baseline fails.
\end{enumerate}

The remainder of this paper is organized as follows.
Section~\ref{sec:plans_overview} provides an overview of the proposed framework.
Section~\ref{sec:plans_label_generation} describes the automatic generation of point labels and training weights using vision-assisted multi-reference DGNSS.
Section~\ref{sec:plans_point_supervision} presents the point-supervised learning method, its implementation with different visual models, and the use of predicted fault probabilities for GNSS observation weighting.
Section~\ref{sec:plans_experiments} presents the experimental setup and analyzes GNSS residuals, predicted fault probability maps, and positioning performance.
Finally, Section~\ref{sec:plans_conclusion} summarizes the main findings and outlines future work.

\section{Problem Formulation and Framework Overview}
\label{sec:plans_overview}

\begin{figure*}[t]
    \centering
    \includegraphics[width=\textwidth]{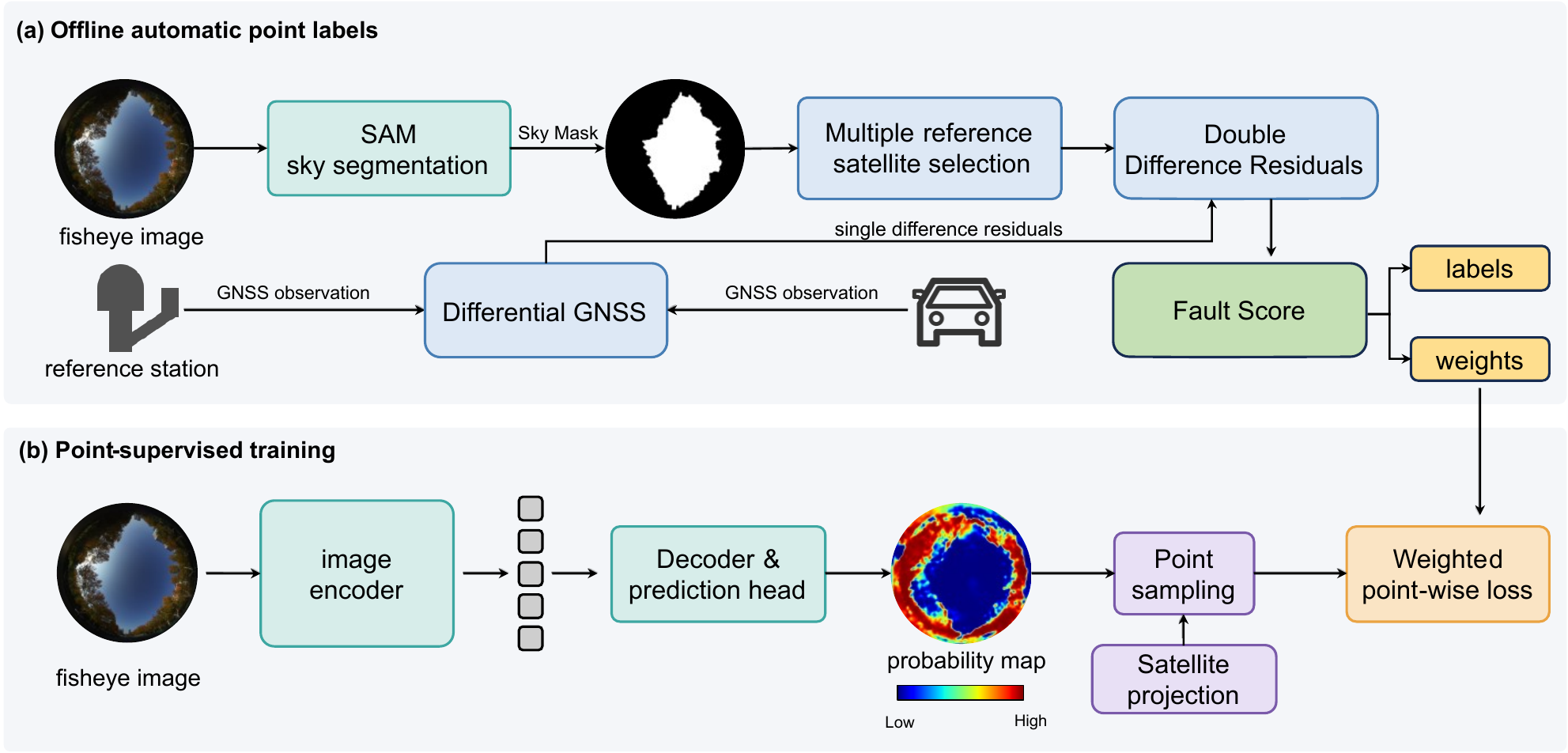}
    \caption{Overview of the proposed point-supervised framework.
    (a) Offline automatic point-label generation using sky segmentation,
    multiple reference satellite selection, and double-difference residuals.
    (b) Point-supervised visual learning using the generated labels and
    residual-based sample weights through a weighted point-wise loss.}
    \label{fig:framework}
\end{figure*}

The proposed framework uses point supervision to predict GNSS observation faults from fisheye sky images, with the aim of improving positioning accuracy in urban environments. As illustrated in Fig.~\ref{fig:framework}, it comprises the offline automatic point label generation and network training with point supervision. The first stage derives supervision from GNSS observations, while the second uses this supervision to learn the relationship between visual features and observation faults.

In the offline stage, SAM generates a sky mask to guide the selection of multiple reference satellites within each constellation. GNSS observations collected at the reference station and the vehicle are used to compute single difference residuals. For each target satellite, double-difference residuals are then constructed with respect to the selected reference satellites and aggregated into an observation fault score. The resulting scores are converted into point labels and corresponding training weights, which are associated with the satellite projection locations in the fisheye image.

During training, the network predicts a dense fault probability map from each fisheye image. 
The predicted probabilities are sampled at satellite projection locations to compute a weighted pointwise loss using the corresponding fault labels and sample weights. 
This design enables the network to estimate fault probabilities for GNSS observations using automatically generated sparse supervision, without requiring dense pixelwise fault annotations.

\section{Automatic Point Label Generation Using Vision-Assisted Multi Reference DGNSS}
\label{sec:plans_label_generation}

\subsection{Principles of DGNSS}
\label{sec:plans_dgnss}
To achieve this, measurement errors caused by multipath and NLOS propagation should be distinguished from those arising from other sources. In the GNSS field, we exploit short baseline DGNSS observations and pseudorange double differencing to suppress common GNSS errors and thereby isolate the measurement deviations associated with local environmental obstructions. The raw GNSS pseudorange observation can be expressed as
\begin{equation} 
P_{r}^{s} = \rho_{r}^{s} + c(dt_r - dt^s) + I_{r}^{s} + T_{r}^{s} + \epsilon_{MP/NLOS}^{s} + \epsilon_{r}^s
\end{equation}
where $\rho_{r}^{s}=\|\mathbf x_r-\mathbf x^{\,s}\|$ denotes the geometric range between receiver $r$ and satellite $s$ in the ECEF frame. The terms $dt_{r}$ and $dt^{s}$ denote the receiver and satellite clock offsets, respectively, while $I_{r}^{s}$ and $T_{r}^{s}$ denote the ionospheric and tropospheric delays. The term $\epsilon_{MP/NLOS}^{s}$ represents the additional bias induced by NLOS or multipath, and $\epsilon_{r}^s$ is the measurement noise, which is modeled as zero mean Gaussian noise.

\begin{figure}
    \centering
    \includegraphics[width=0.75\linewidth]{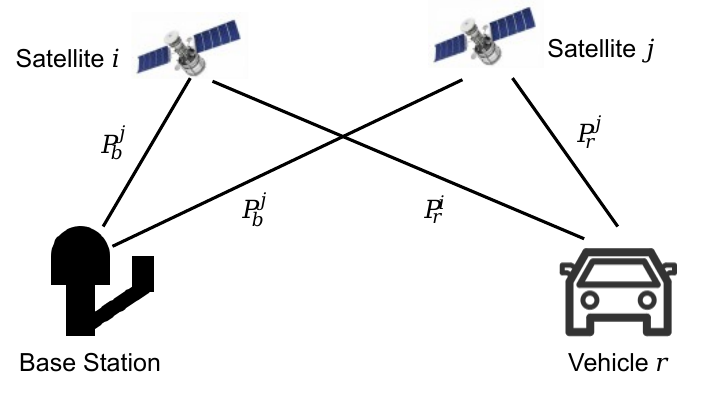}
    \caption{Schematic of the DGNSS principle}
    \label{fig:dgnss}
\end{figure}

DGNSS is a technique that is used to effectively eliminate ionospheric and tropospheric delays. As illustrated in Fig.~\ref{fig:dgnss}, this approach relies on a nearby reference station in the open sky without NLOS and multipath, whose position is accurately known. When the baseline between the rover and reference station is sufficiently short, for example approximately 10 km as described in~\cite{kaplan2017understanding}, the atmospheric conditions experienced by the two receivers can be regarded as highly correlated. Under this short-baseline assumption, the ionospheric and tropospheric components are largely common-mode and can therefore be effectively removed through pseudorange double differencing. 

First, the Single Difference (SD) between receivers is formed by subtracting the measurements of the base station $b$ from those of the rover $r$ for the same satellite $k$. The formula is as follows
\begin{equation} 
    \Delta P_{rb}^{k} = \Delta \rho_{rb}^{k} + c\Delta dt_{rb} + \Delta\epsilon_{MP/NLOS}^{k} + \Delta \epsilon_{rb}^{k} 
\end{equation}
where $\Delta \rho_{rb}^{k}=\rho_{r}^{k}-\rho_{b}^{k}$ denotes the single-differenced geometric range, and $c\Delta dt_{rb}$ represents the relative receiver clock bias between the rover and the base station. Since the reference station is located in an ideal open-sky environment, its multipath error is negligible. Thus, the differential term $\Delta \epsilon_{MP/NLOS}^{k}$ is dominated by the multipath error from the rover. 

Subsequently, a Double Difference (DD) is formed by differencing the SD observations associated with the current satellite $k$ and a reference satellite $j$. This step removes the receiver clock bias common to both observations. The DD observation can be written as
\begin{equation} 
\nabla \Delta P_{rb}^{jk} = \nabla \Delta \rho_{rb}^{jk} + \nabla \Delta \epsilon_{MP/NLOS}^{jk} + \nabla \Delta \epsilon_{rb}^{jk}
\end{equation}
where
\begin{equation} 
\nabla \Delta \rho_{rb}^{jk} = \Delta \rho_{rb}^{k} - \Delta \rho_{rb}^{j}
\end{equation}
denotes the double difference geometric range. By eliminating clock biases and atmospheric delays, the residual term $\nabla \Delta \epsilon_{MP/NLOS}^{jk}$ becomes a direct indicator of signal quality. The term $\nabla \Delta \epsilon_{rb}^{jk}$ denotes measurement noise, which is commonly modeled as zero-mean Gaussian white noise~\cite{Teunissen1995}.

To characterize the observation error of satellite \(k\), the DD pseudorange residual is defined as

\begin{equation}
\begin{aligned}
\varepsilon_{rb}^{jk}
&= \nabla \Delta P_{rb}^{jk}
- \nabla \Delta \rho_{rb}^{jk} \\
&= \nabla \Delta \epsilon_{\mathrm{MP/NLOS}}^{jk}
+ \nabla \Delta \epsilon_{rb}^{jk}.
\end{aligned}
\end{equation}

When the reference satellite $j$ is selected from reliable LOS observations, its propagation error at the rover is assumed to be small. 
Under this condition, the DD residual $\varepsilon_{rb}^{jk}$ primarily reflects the propagation error associated with satellite $k$, together with the remaining measurement noise. 
For LOS signals with negligible multipath effects, double-difference pseudorange residuals are expected to remain relatively small. 
In contrast, multipath and NLOS propagation can introduce additional pseudorange biases, potentially resulting in larger residual magnitudes.
\subsection{SAM-Assisted Multi-Reference Residual Estimation}

However, the reliability of the double difference residuals depends critically on the quality of the reference satellite $j$. A conventional strategy is to select the satellite with the highest elevation angle as the reference, based on the assumption that a high elevation satellite is more likely to have an unobstructed propagation path. In deep urban canyons, however, this assumption may not hold, since even high elevation satellites can be obstructed by surrounding buildings. 
An unreliable reference satellite may therefore propagate its measurement errors into all double difference observations formed using that satellite.

Specifically, a semantic segmentation network
(e.g., SAM or U-Net) is employed to process fisheye images and generate
a binary sky mask, distinguishing the open sky from environmental
obstacles:
\begin{equation}
\operatorname{sky}_{t}(u,v)=
\begin{cases}
1, & \text{sky},\\
0, & \text{otherwise}.
\end{cases}
\label{eq:sky_mask}
\end{equation}

For each constellation $g$, let $(u_s,v_s)$ denote the projected pixel coordinates of satellite $s$ in the fisheye image. Based on the sky mask, satellites projected onto the sky region are identified as LOS candidates. Among these candidates, only those with elevation angles above $45^\circ$ are retained to form the reference satellite set:

\begin{equation}
S_{\mathrm{ref},t}^{(g)}
=
\left\{
s \in \mathcal{S}_{t}^{(g)}
\;\middle|\;
\operatorname{sky}_{t}(u_{s},v_{s})=1,\;
el_{s,t}>45^\circ
\right\}.
\label{eq:reference_satellite_selection}
\end{equation}
where $\mathcal{S}_{t}^{(g)}$ denotes the set of observed satellites in constellation $g$ at epoch $t$, and $el_{s,t}$ represents the elevation angle of satellite $s$. Instead of using a single reference satellite, all selected satellites in $\mathcal{S}_{\mathrm{ref},t}^{(g)}$ are jointly used to construct double-difference pseudorange residuals within each constellation.

Accurate computation of the double difference residual also requires precise knowledge of the rover position during label generation. Let $\mathbf{x}_{gt}$ denote the ground truth position of the rover. By fixing the rover coordinates to $\mathbf{x}_{gt}$, the geometric ranges and their corresponding double differences can be computed deterministically rather than estimated from a potentially degraded positioning solution. 
For satellite $k$ in constellation $g$, the residual is calculated by averaging the double difference pseudorange residuals obtained using the selected reference satellites within the same constellation:
\begin{equation}
e_{t}^{k,(g)}
=
\frac{1}{M_{t}^{(g)}}
\sum_{\substack{
j \in S_{\mathrm{ref},t}^{(g)}
}}
\left(
\nabla\Delta P_{rb,t}^{jk}
-
\nabla\Delta \rho_{rb,t}^{jk}
\right).
\end{equation}
where $M_{t}^{(g)}=|S_{\mathrm{ref},t}^{(g)}|$ denotes the number of available reference satellites. Since satellite visibility and environmental conditions vary over time, $M_{t}^{(g)}$ may change from epoch to epoch. The resulting residual $e_t^{k,(g)}$ primarily reflects the propagation error associated with satellite $k$, while also containing residual measurement noise and other unmodeled errors.

\subsection{Residual-Derived Point Labels and Importance Weights}
\label{sec:plans_labels_weights}

The resulting residual magnitudes reflect the extent to which GNSS observations are affected by multipath and NLOS propagation, providing a basis for generating binary point labels with corresponding sample importance weights.

For a target satellite $s$ with a valid residual at epoch $t$, the binary point label is defined as
\begin{equation}
    y_t^s =
    \begin{cases}
        1, & |e_t^s| > \tau,\\
        0, & |e_t^s| \leq \tau.
    \end{cases}
    \label{eq:plans_fault_label}
\end{equation}
The threshold $\tau$ is a residual threshold in range units.


Under normal conditions, healthy observations typically dominate the
sparse labels. However, undetected faulty observations with large
pseudorange errors can bias the initial position estimate obtained by
weighted least squares (WLS), thereby affecting the residuals used for
subsequent quality control. To place greater emphasis on identifying
these faults during training, we use the magnitude of the DGNSS
double-difference residual as an indicator of observation error
severity and define the training importance weight for satellite
$k$ at epoch $t$ as
\begin{equation}
    w_t^k = \alpha_{y_t^k}
    \ln\left(1+\frac{\min(|e_t^k|,r_{\mathrm{cap}})}{r_0}\right)
    \label{eq:plans_importance_weight}
\end{equation}
where $\alpha_{y_t^k}$ is a coefficient determined by the observation label, with $\alpha_0=0.2$ and $\alpha_1=1.6$ assigned to healthy and faulty observations, respectively, in this study. 
The parameter $r_0>0$ is a reference residual scale that makes the logarithmic argument dimensionless.
The larger coefficient assigned to faulty observations increases the loss penalty for missed detections, while the logarithmic term gives greater importance to observations with larger residual magnitudes up to the clipping limit $r_{\mathrm{cap}}$. 
These weights are applied only to the training loss and are distinct from the observation weights used in WLS. This design encourages the model to identify severe faults so that their influence on the initial positioning solution can be suppressed.

It should be noted that double-difference pseudorange residuals can vary widely across satellite observations within the same epoch. Although the logarithmic mapping compresses these differences, exceptionally large residuals may still contribute disproportionately to the sum used for weight normalization, leaving the remaining observations with very small normalized weights and limited contributions to the training loss. 
We therefore set \(r_{\mathrm{cap}}=20\) m to prevent residuals beyond this magnitude from further increasing the component of the importance weights that depends on the residual magnitude. 

\section{Point-Supervised Visual GNSS Observation Fault Detection}
\label{sec:plans_point_supervision}

\subsection{Dense Fault-Probability Map Prediction}
\label{sec:plans_dense_prediction}

Given an image $I_t$, the model predicts a dense fault probability
map:
\begin{equation}
    \hat P_t = f_\theta^{(m)}(I_t),
    \label{eq:plans_dense_map}
\end{equation}
where $m$ denotes the model implementation and $\theta$ its trainable parameters. At each output location $(u,v)$,
$\hat P_t(u,v) \in [0,1]$ represents the predicted probability
that a GNSS observation from the corresponding viewing
direction is faulty.

For target satellite $k$, its direction is projected onto
the image plane. Let $\mathbf p_t^k = (u_t^k,v_t^k)$ denote the resulting coordinate. The predicted fault probability of the
corresponding GNSS observation is then obtained by sampling the probability map:
\begin{equation}
    \hat q_t^k = \hat P_t(\mathbf p_t^k).
    \label{eq:plans_satellite_fault_prob}
\end{equation}

\subsection{Sparse Point Supervision and Residual-Weighted Loss}
\label{sec:plans_loss}

\textbf{Sparse Supervision Strategy.}
At epoch $t$, let $\mathcal V_t$ denote the indices of satellite
observations with valid labels and projected locations within
the output grid. The supervision set is
\begin{equation}
    \mathcal S_t =
    \left\{
        (\mathbf p_t^k, y_t^k, w_t^k)
        : k \in \mathcal V_t
    \right\}.
    \label{eq:plans_supervision_set}
\end{equation}

Although the model predicts a dense map, supervision is restricted to the satellite observations indexed by $\mathcal V_t$. 
Unlabeled image locations do not contribute to the loss.

\textbf{Residual-Weighted Loss Function.}
The importance weights $w_t^k$ account for class imbalance
and give greater emphasis to observations with larger
residual magnitudes. The weighted binary cross entropy
loss at epoch $t$ is defined as
\begin{equation}
    \mathcal L_t =
    -\frac{
        \displaystyle\sum_{k\in\mathcal V_t}
        w_t^k
        \left[
            y_t^k \log \hat q_t^k
            + (1-y_t^k)\log(1-\hat q_t^k)
        \right]
    }{
        \displaystyle\sum_{k\in\mathcal V_t} w_t^k
        + \epsilon
    },
    \label{eq:plans_point_loss}
\end{equation}
where $\hat q_t^k = \hat P_t(\mathbf p_t^k)$ is the
predicted fault probability for observation $k$.
The loss compares this probability with the binary
label $y_t^k$ and is normalized by the total importance
weight of the valid observations. A small constant
$\epsilon>0$ is added to prevent division by zero.
The same objective is used to train all model
implementations.


\begin{table*}[htbp]
    \centering
    \caption{Model configurations and training strategies for point-supervised GNSS observation fault prediction.}
    \label{tab:model_configurations}

    \small
    \setlength{\tabcolsep}{1.5pt}
    \renewcommand{\arraystretch}{1.0}
    \renewcommand{\tabularxcolumn}[1]{m{#1}}

\begin{tabularx}{\textwidth}{
    @{}
    >{\raggedright\arraybackslash}m{0.09\textwidth}
    >{\raggedright\arraybackslash}m{0.25\textwidth}
    >{\raggedright\arraybackslash}X
    >{\raggedright\arraybackslash}m{0.23\textwidth}
    @{}
}
        \toprule
        \textbf{Model} &
        \textbf{Visual encoder} &
        \textbf{Decoding and prediction modules} &
        \textbf{Training strategy} \\
        \midrule

        U-Net &
        ImageNet-pretrained ResNet-34 &
        Progressive upsampling decoder with skip connections and a convolutional prediction head &
        Full fine tuning. \\
        \addlinespace[5pt]

        DeepLabv3+ &
        ImageNet-pretrained ResNet-34 &
        Atrous spatial pyramid pooling (ASPP), shallow feature fusion, and a convolutional prediction head &
        Full fine tuning. \\
        \addlinespace[5pt]

        DINOv3 &
        DINOv3 ViT-S/16 with self-supervised pretraining &
        Fusion of intermediate Transformer features followed by progressive upsampling and a convolutional prediction head &
        Full fine tuning. \\
        \addlinespace[5pt]

        SAM3 &
        Pretrained SAM3 Vision Transformer backbone &
        Convolutional prediction head followed by bilinear upsampling to the target output resolution &
        Frozen encoder training only. \\

        \bottomrule
    \end{tabularx}

\end{table*}

\begin{figure*}[htbp]
    \centering
    \includegraphics[width=0.9\linewidth]{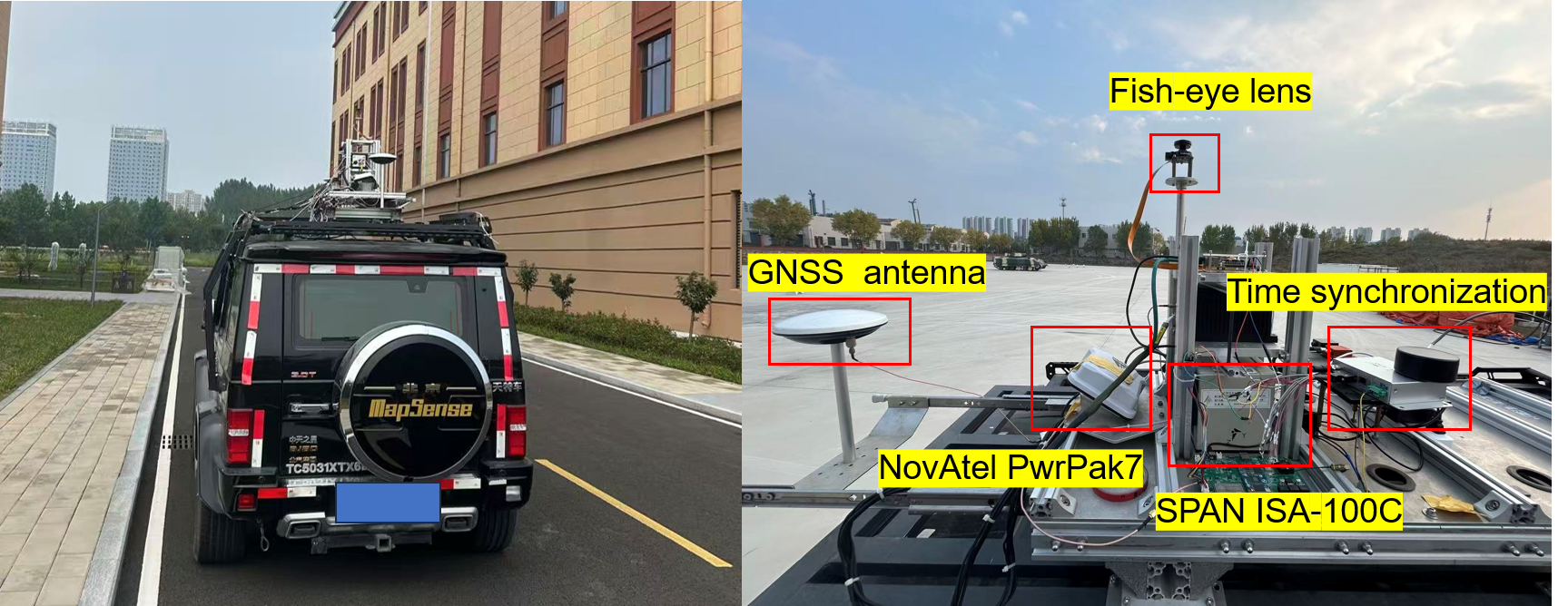}
    \caption{Experimental platform and sensor configuration. The left panel shows the vehicle-mounted setup, while the right panel illustrates the specific layout of the GNSS antenna, fisheye vision module, and the reference navigation system on the roof rack.}
    \label{fig:hardware_setup}
\end{figure*}

\subsection{Instantiation with Different Visual Models}
\label{sec:plans_models}

We further evaluate the effectiveness of the proposed point-supervised framework for GNSS observation fault prediction and its applicability across different network architectures and pretrained representations.

To achieve this target, we implement four dense prediction models based on U-Net~\cite{Ronneberger2015UNet}, DeepLabv3+~\cite{Chen2018DeepLabv3Plus}, DINOv3~\cite{Simeoni2025DINOv3}, and SAM3~\cite{Carion2025SAM3}. All four configurations share a common point-level supervision scheme based on the label generation method and weighted loss function described above. Table~\ref{tab:model_configurations} summarizes the visual encoders, decoding and prediction modules, and training strategies used in each configuration.

\textbf{U-Net} fuses features from corresponding encoder and decoder stages through skip connections, combining local spatial details with deeper contextual information during progressive upsampling.
We adopt a ResNet-34-based U-Net as the baseline to assess whether sparse point-level supervision enables the model to exploit local cues near sky obstruction boundaries, including building contours, canopy edges, and gaps in vegetation, for predicting GNSS observation anomalies at satellite projection locations.

\textbf{DeepLabv3+} uses atrous spatial pyramid pooling (ASPP) to aggregate contextual information over multiple receptive fields and incorporates low-level spatial features through its decoder to refine boundary details. Given the variations in the apparent size and spatial distribution of obstructions in fisheye images, this model is included to evaluate the proposed framework with a representation that combines multiscale context and local spatial information.

\textbf{DINOv3} provides transferable visual representations learned through self supervision during pretraining on large datasets. Its visual Transformer encoder supports interactions across image regions, and features extracted at different Transformer depths are integrated through a dense prediction decoder. This configuration is used to assess whether sparse labels
at satellite projection locations can adapt generic pretrained representations to GNSS observation fault prediction without additional manual pixel-level annotations.

\textbf{SAM3} provides pretrained visual representations learned from detection and segmentation tasks.
We retain its pretrained visual encoder as a frozen feature
extractor and train a new dense prediction head using the
proposed point-level loss. This configuration evaluates the transferability of segmentation-related structural representations to GNSS observation fault prediction when task adaptation is restricted to the prediction head.

\subsection{Satellite Observation Fault Probability Prediction}

For GNSS positioning, the predicted fault probability
\(\hat q_t^k\) is converted into a visual weight for the
corresponding satellite observation:
\begin{equation}
    w_{t,\mathrm{vis}}^k = 1-\hat q_t^k,
    \label{eq:plans_visual_weight}
\end{equation}
so that observations with higher predicted fault probabilities
receive lower weights. At epoch \(t\), the visual weights of the
\(N_t\) available satellite observations are assembled into the
weighting matrix
\begin{equation}
    \mathbf{W}_t
    =
    \operatorname{diag}
    \left(
        w_{t,\mathrm{vis}}^1,\ldots,
        w_{t,\mathrm{vis}}^{N_t}
    \right).
    \label{eq:plans_visual_weight_matrix}
\end{equation}

For WLS positioning, \(\mathbf{z}_t\) denotes the differences between the corrected pseudorange measurements and the pseudoranges computed from the current receiver state estimate, while \(\mathbf{H}_t\) is the corresponding geometry matrix.
The linearized observation model takes the form \cite{Li2023MLGNSS}
\begin{equation}
    \mathbf{z}_t
    =
    \mathbf{H}_t\,\Delta\mathbf{x}_t
    +
    \boldsymbol{\varepsilon}_t,
    \label{eq:plans_linearized_observation}
\end{equation}
where \(\Delta\mathbf{x}_t\) contains the corrections to the receiver position and clock bias parameters, and \(\boldsymbol{\varepsilon}_t\) denotes the observation error vector. 
Incorporating the visual weighting matrix, the state correction is estimated as
\begin{equation}
    \Delta\hat{\mathbf{x}}_t
    =
    \left(
        \mathbf{H}_t^{\mathsf T}
        \mathbf{W}_t
        \mathbf{H}_t
    \right)^{-1}
    \mathbf{H}_t^{\mathsf T}
    \mathbf{W}_t
    \mathbf{z}_t.
    \label{eq:plans_visual_wls_solution}
\end{equation}
The receiver state is updated by adding the estimated correction,
with \(\mathbf{z}_t\) and \(\mathbf{H}_t\) recomputed at each
iteration. The visual weights remain fixed throughout the
iterations within each epoch, controlling the contribution of
each observation to the positioning solution.

\section{Experimental Material and Evaluation Protocol}
\label{sec:plans_experiments}

\begin{figure*}[htbp]
    \centering
    \includegraphics[width=0.9\linewidth]{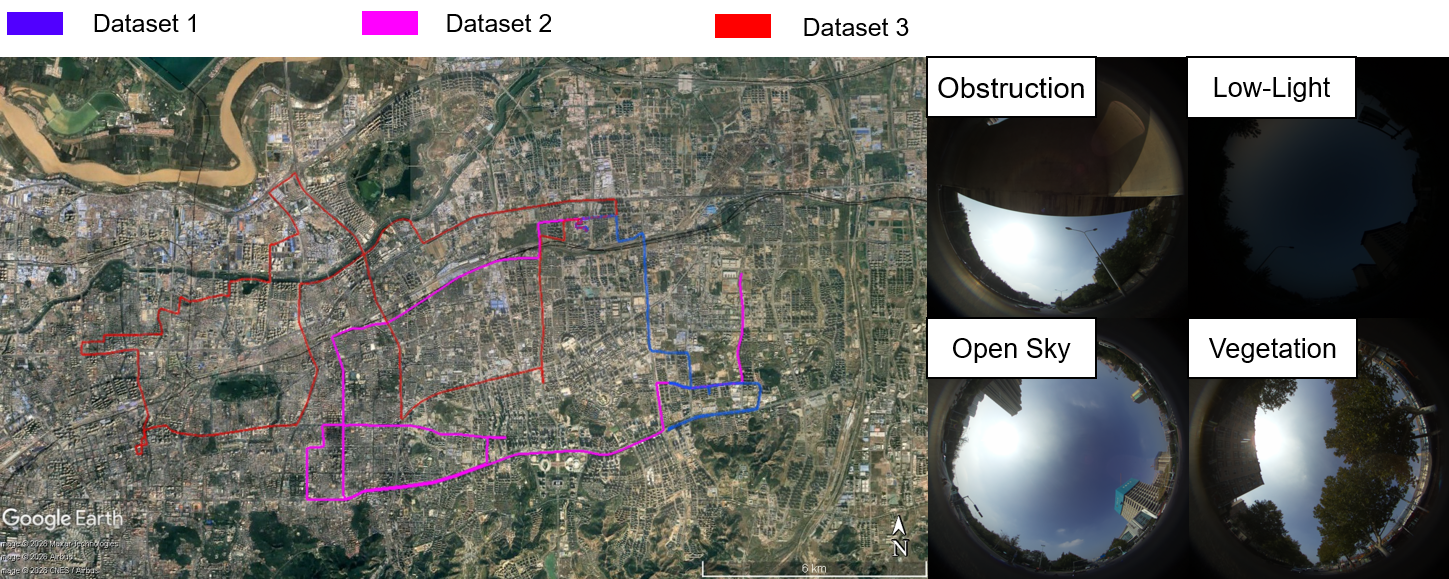}
    \caption{Trajectories of the experimental datasets and representative environmental examples. The left panel shows the three data routes with color-coded trajectories (Dataset~1, Dataset~2, Dataset~3). The four right panels present fisheye images illustrating typical local conditions.}
    \label{fig:trajectory of datasets}
\end{figure*}

This section evaluates the proposed point supervised framework using real world data collected in complex urban environments. We first analyze the characteristics of the collected dataset and evaluate the satellite fault prediction performance across different visual architectures to examine the effectiveness and generalizability of the proposed framework. The resulting fault probabilities are subsequently incorporated into GNSS positioning to assess their contribution to positioning accuracy.

\subsection{Experimental Setup}

This experiment was conducted using a vehicle-mounted multi-sensor integrated navigation platform. The hardware architecture is shown in Fig.~\ref{fig:hardware_setup}. For raw GNSS observation acquisition, a NovAtel PwrPak7 measurement-grade receiver was used at a sampling frequency of 1 Hz. The receiver tracked dual-frequency signals from multiple constellations, including GPS (L1/L2), BeiDou (B1I/B2I), and Galileo (E1/E5b). To characterize the surrounding environment, the platform was also equipped with a vision module consisting of a Raspberry Pi 5 and a CHANCCTV CH316B fisheye lens. This lens provides an ultra-wide field of view (FOV) of $185^{\circ}$, enabling continuous sky-view imaging, and the image stream was strictly synchronized with the GNSS time reference. To evaluate positioning performance, centimeter-level ground-truth trajectories were generated using a SPAN ISA-100C system through post-processed tightly coupled GNSS/INS integration.

The experimental data were collected in a representative urban environment in Jinan, China. Fig.~\ref{fig:trajectory of datasets} shows the vehicle trajectories of the three selected datasets (Dataset 1--Dataset 3) projected on Google Earth. The test routes covered challenging urban scenarios, including tree-lined streets, dense urban canyons with tall buildings, underpass sections, and low-light conditions. These environments introduce severe signal blockage and pronounced multipath effects.

\subsection{Dataset Composition and GNSS Residual Analysis}

To clarify the composition of the experimental data samples, we first calculated the proportion of satellite observations in different environmental categories. Specifically, based on the environmental category corresponding to the projection position of the satellite in the fisheye image, the observations were classified into three categories: sky, trees, and buildings. The statistical results are shown in Table~\ref{tab:dataset_semantic_distribution}, providing a data basis for the subsequent analysis of the relationship between environmental categories and GNSS observation quality.

\begin{table}[htpb]
    \centering
    \caption{ Visual Semantic Composition of the Three Datasets.}
    \label{tab:dataset_semantic_distribution}

    \setlength{\tabcolsep}{8pt}
    \renewcommand{\arraystretch}{1.1}

    \resizebox{0.9\columnwidth}{!}{
    \begin{tabular}{lccc}
        \hline
        Dataset & Sky (\%) & Tree (\%) & Building (\%) \\
        \hline
        Dataset 1 & 79.04 & 4.67 & 16.29 \\
        Dataset 2 & 68.88 & 9.88 & 21.24 \\
        Dataset 3 & 57.71 & 12.89 & 29.40 \\
        \hline
    \end{tabular}
    }
\end{table}

As shown in Table~\ref{tab:dataset_semantic_distribution}, observations in the sky category constitute the majority in all three datasets. Dataset 1 has the highest proportion of sky observations, at 79.04\%. In contrast, Dataset 3 has the highest proportions of observations in the tree and building categories, at 12.89\% and 29.40\%, respectively.
In this study, Datasets 1 and 3 are used for model training, while Dataset 2 is used for testing.

\begin{figure}[htpb]
    \centering
    \includegraphics[width=\linewidth]{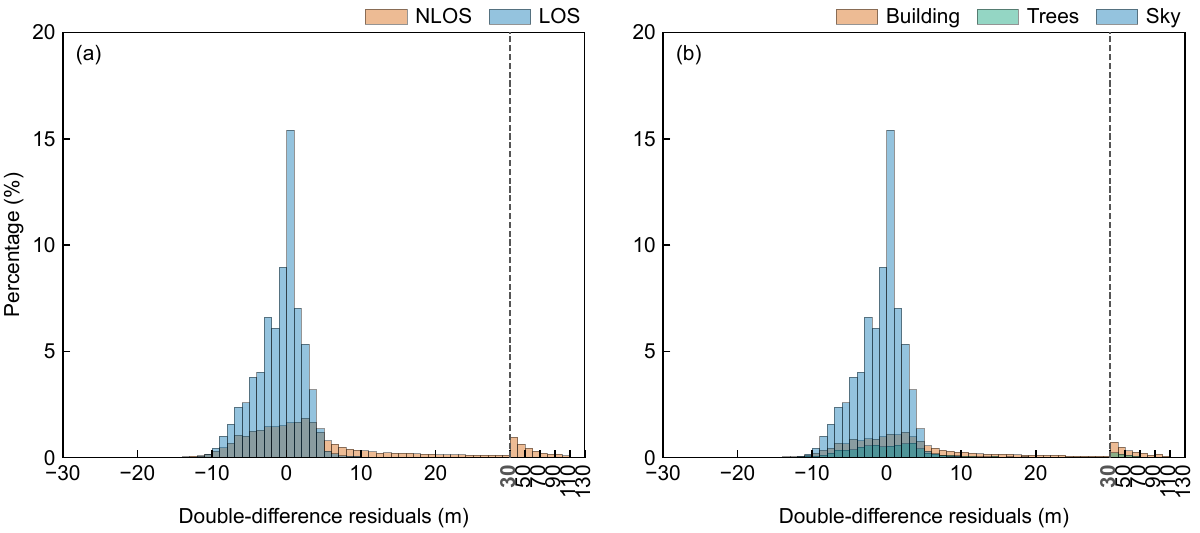}
    \caption{Distributions of double-difference pseudorange residuals in Dataset 2: (a) LOS and NLOS observations; (b) observations in the sky, tree, and building categories.}
    \label{fig:los_nlos_residual_distribution}
\end{figure}

\begin{figure*}[htpb]
    \centering
    \includegraphics[width=\textwidth]{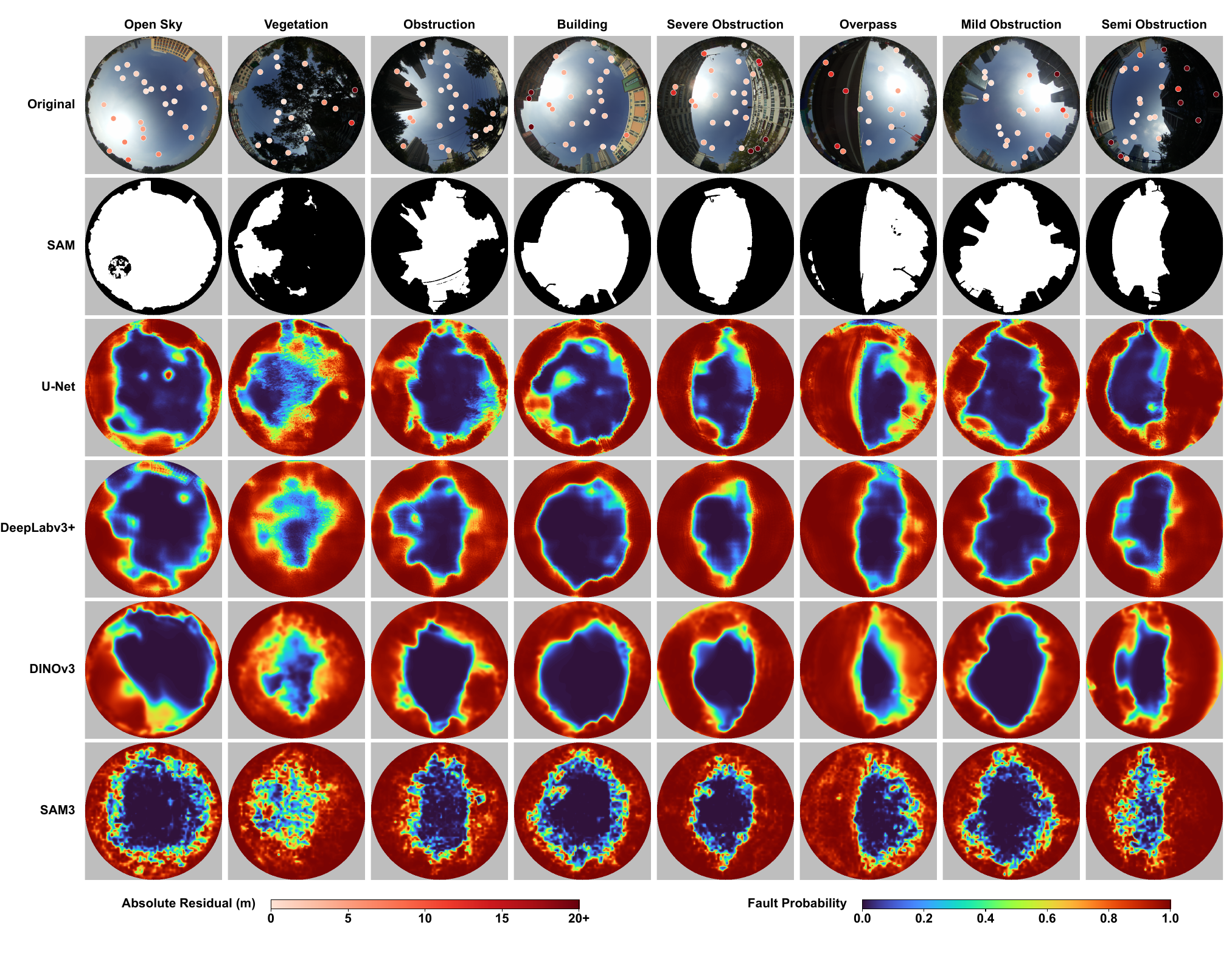}
    \caption{Comparison of fisheye sky images and fault probability
    maps generated by the point-supervised framework across different
    scenes. Satellite marker colors indicate the absolute
    double-difference residuals of the corresponding observations
    (left color bar, m), with $20+$ denoting values of $20$~m or greater.
    The right color bar indicates predicted fault probabilities
    ranging from $0$ to $1$.}
    \label{fig:model_comparison}
\end{figure*}
\begin{figure*}[htpb]
    \centering
    \includegraphics[width=\textwidth]{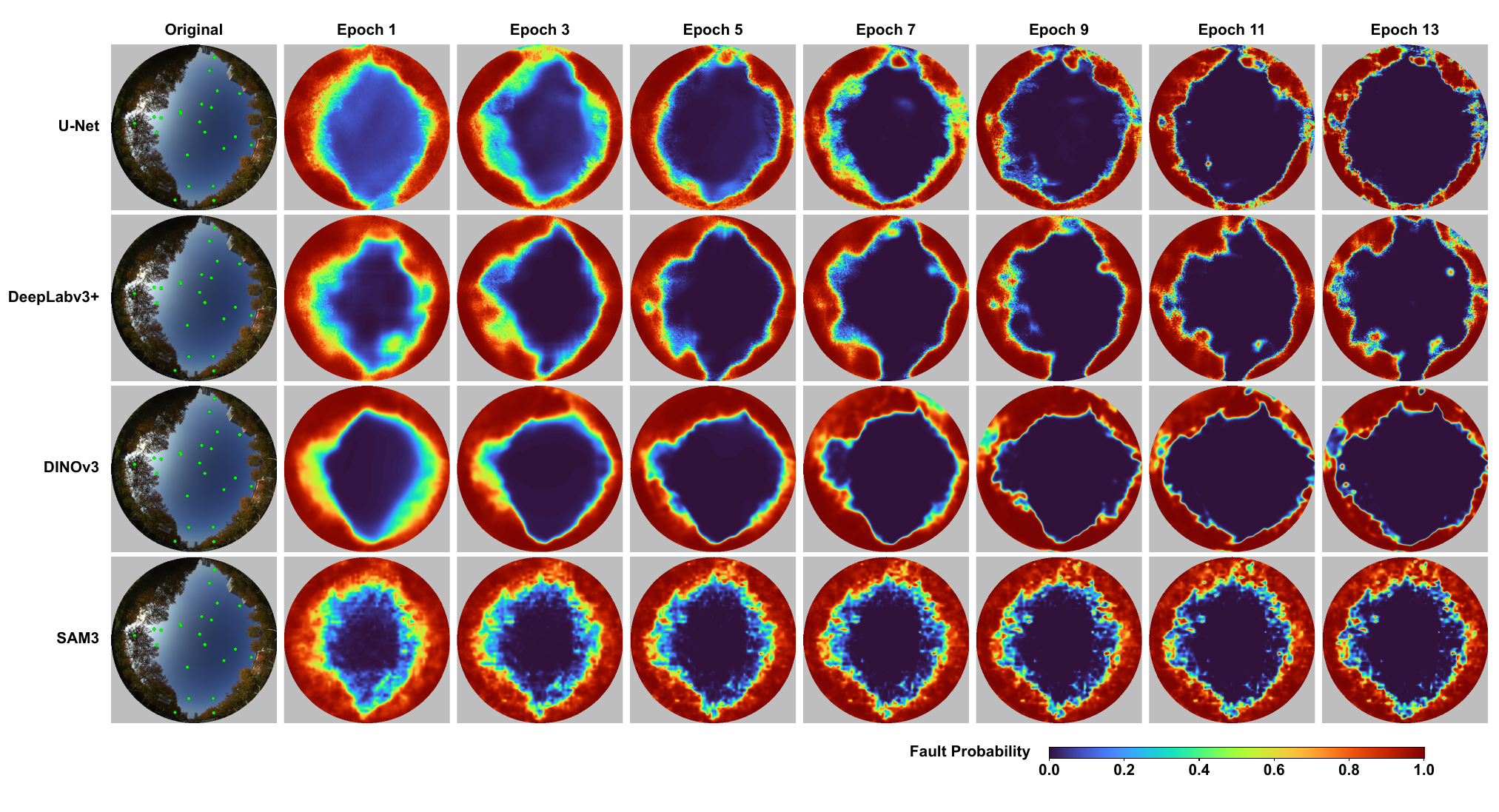}
    \caption{Fault probability maps predicted by four point-supervised
    models for the same fisheye image at different training epochs.
    The first column shows the original image, and the remaining
    columns show predictions at the indicated epochs. The color bar
    indicates predicted fault probabilities ranging from $0$ to $1$.}
    \label{fig:training_epoch_comparison}
\end{figure*}

Fig.~\ref{fig:los_nlos_residual_distribution} compares the double-difference residual distributions of satellite observations in Dataset 2. As shown in Fig.~\ref{fig:los_nlos_residual_distribution}(a), within the range from -30 to 30 m, LOS residuals are concentrated near zero, whereas NLOS residuals are more dispersed and exhibit a more pronounced positive tail. NLOS observations are also more frequent than LOS observations in the range of large absolute residuals. Fig.~\ref{fig:los_nlos_residual_distribution}(b) further compares the residual distributions across the sky, tree, and building categories. Sky residuals are concentrated near zero, and some observations in the tree category also have small residuals. In contrast, observations in the building category exhibit a broader residual distribution and a more positive tail at large residual magnitudes.

Although the overall distributions differ, LOS and NLOS residuals overlap substantially near zero, while some LOS observations also exhibit large residuals. This indicates that LOS/NLOS classification captures general differences in observation quality but cannot reliably distinguish healthy from faulty measurements.
To establish a residual-based labeling criterion, a Gaussian distribution is fitted to the signed residuals of LOS observations in the training data, yielding the estimated mean \(\hat{\mu}\) and variance \(\hat{\sigma}^{2}\). The threshold for the absolute residuals is then defined as \(\tau=|\hat{\mu}|+3\hat{\sigma}\), such that the symmetric interval \([-\tau,\tau]\) encloses the fitted interval \([\hat{\mu}-3\hat{\sigma},\,\hat{\mu}+3\hat{\sigma}]\).
Measurements with absolute residuals exceeding $\tau$ are classified as faulty, whereas the remaining measurements are classified as healthy. These labels reflect measurement reliability rather than signal visibility alone. This motivates the development of a method that extracts information beyond binary sky visibility from fisheye images, which is the central objective of this study.

\subsection{Analysis of Fault Probability Maps}

Although both buildings and trees can cause obstructions in the satellite line-of-sight direction, the corresponding signal propagation conditions for the two are not the same. Signals reflected from building surfaces may also cause multipath in observations containing a direct signal component. 
Therefore, visual visibility cannot fully reflect the reliability of GNSS observations. 
To examine whether the point supervision framework can provide observation quality information beyond visibility, Fig.~\ref{fig:model_comparison} compares sky masks generated by SAM with fault probability maps predicted by four point-supervised models under different obstruction conditions. The color of the satellite markers represents the absolute value of double-difference residuals of the corresponding observations. 
The SAM masks distinguish only between sky and non-sky regions, whereas point supervision framework uses the residual fault labels at the satellite projection positions as supervision information and outputs a fault probability in the range of  [0,1]. 

In the vegetation scene, SAM classifies most of the tree canopy in the center of the image as non-sky. However, some observations corresponding to satellite projections in this region have relatively small absolute double-difference residuals, whereas observations with larger absolute residuals are present on the right side of the image. Correspondingly, U-Net, DeepLabv3+, and DINOv3 predict relatively low fault probabilities in the central region and generally higher probabilities at the projection locations of observations with larger residuals on the right. This indicates that the point supervision framework does not uniformly consider the tree covered areas as high risk, but forms different risk estimations within the same non-sky category. Some of these predictions are consistent with the observed differences in residual magnitude.

In the open sky scene, the point-supervised models predict low fault probabilities mainly in sky regions, which contain the projection locations of most observations with small absolute double-difference residuals. In the building obstruction and overpass scenes, low fault probabilities are mainly concentrated in visible sky regions, whereas regions occupied by buildings and bridge structures generally have higher predicted probabilities. 
For example, many observations with large residual magnitudes appear on the left side of the overpass scene and on the right side of the scene in the last column of Fig.~\ref{fig:model_comparison}. These examples show that some high-score regions coincide with satellite projection locations associated with large residual magnitudes.

To further examine how the fault probability maps change during training, Fig.~\ref{fig:training_epoch_comparison} presents the predictions of the four models for the same fisheye image at different training epochs. Early in training, all four models generally assign low fault probabilities to the main sky region and high probabilities to the surrounding obstructed regions. As training progresses, the transition regions between low and high probabilities gradually narrow for U-Net, DeepLabv3+, and DINOv3. In contrast, SAM3 continues to produce many small, interspersed patches of high and low probabilities along obstruction boundaries. In the later training stage, the predicted probabilities further become increasingly concentrated near 0 and 1, and the distinction between high and low probability areas becomes more obvious.

\subsection{Positioning Evaluation}
\label{sec:plans_positioning_evaluation}

\begin{figure*}[htbp]
    \centering
    \includegraphics[width=\textwidth]{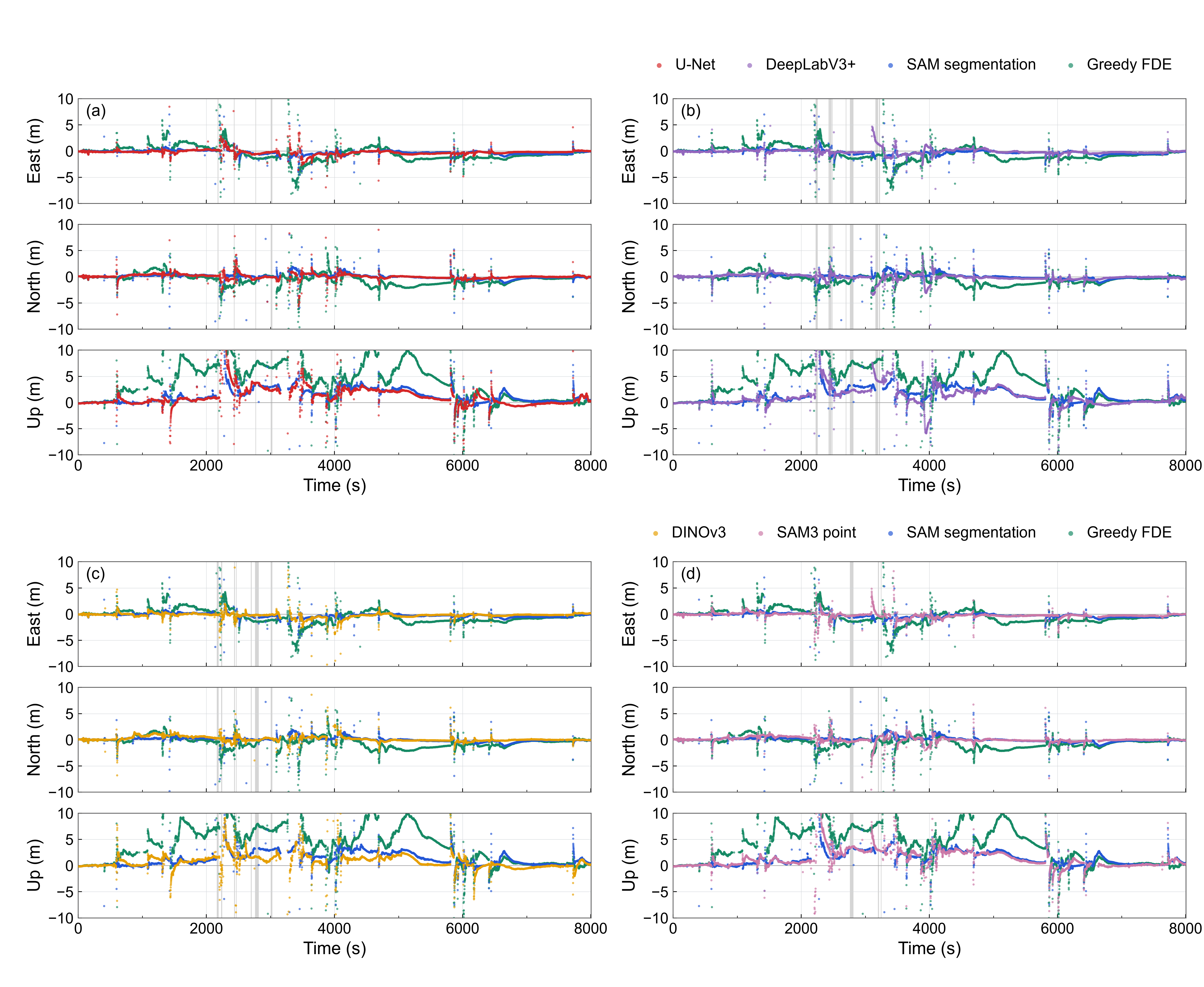}
    \caption{East, north, and up positioning errors on Dataset 2. Each panel compares a point supervision model with SAM binary segmentation and Greedy FDE: (a) U-Net, (b) DeepLabv3+, (c) DINOv3, and (d) SAM3. Gray regions indicate epochs in which SAM binary segmentation fails to provide a positioning solution, whereas the corresponding point supervision model still yields a valid solution.}
    \label{fig:enu}
\end{figure*}

\begin{figure}[htbp]
    \centering
    \includegraphics[width=\linewidth]{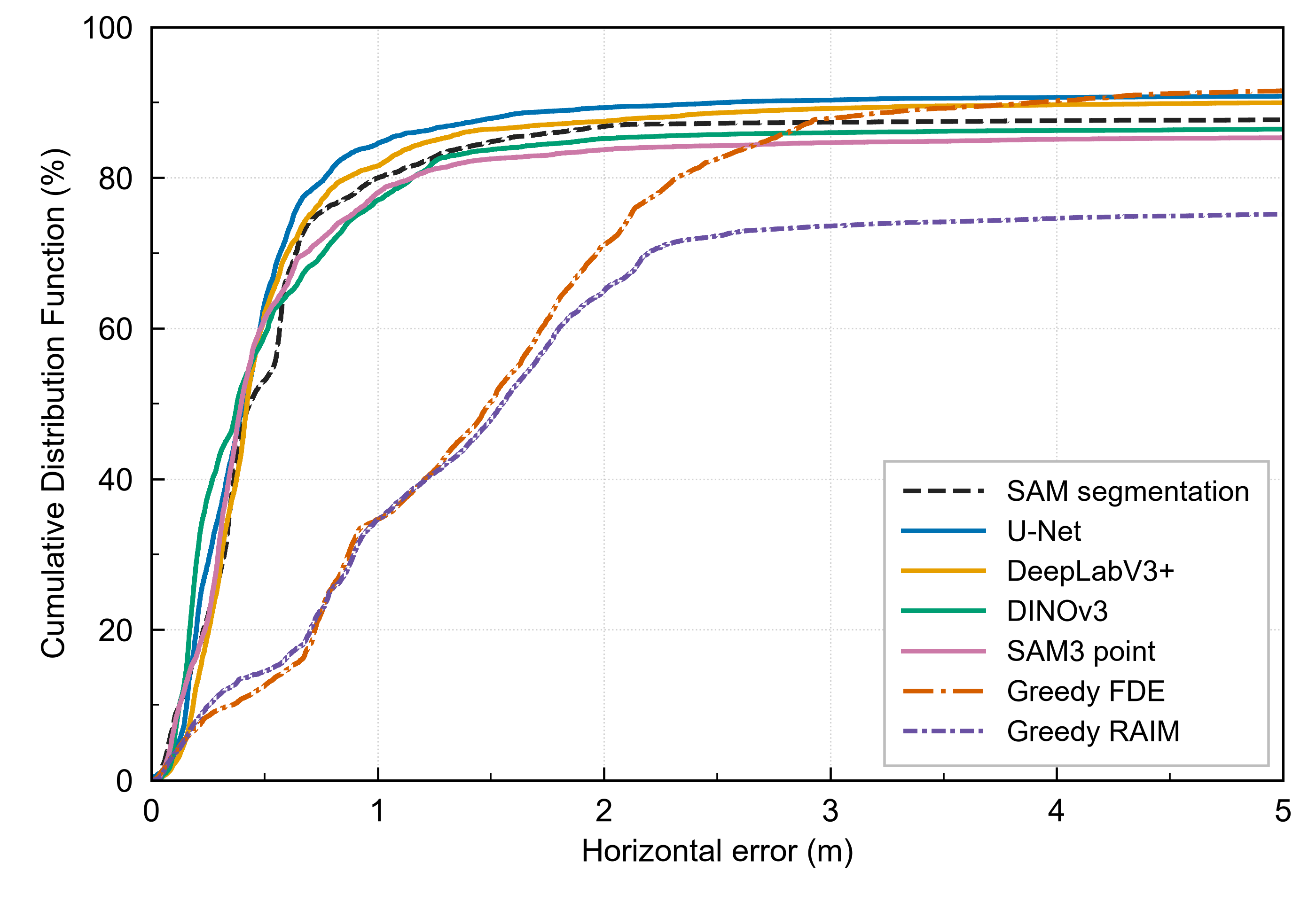}
    \caption{Cumulative horizontal positioning error distributions for the four point-supervised models, SAM-based binary segmentation, Greedy FDE, and Greedy RAIM.}
    \label{fig:cdf}
\end{figure}

Fig.~\ref{fig:enu} compares the east, north, and vertical positioning errors obtained using U-Net, DeepLabv3+, DINOv3, and SAM3 within the point-supervised framework with those obtained using SAM-based binary segmentation and Greedy FDE on Dataset 2. 
As shown in Fig.~\ref{fig:enu}(a)-(d), Greedy FDE exhibits large positioning errors during several intervals. In contrast, SAM-based segmentation and the point-supervised models generally provide more accurate position estimates.
The gray regions indicate epochs at which SAM segmentation yields no positioning solution, whereas the corresponding point-supervised model still provides a valid solution. These results demonstrate that the point-supervised framework can maintain solution availability during some periods when the SAM-based approach cannot provide positioning solutions.

\begin{table}[htbp]
\centering
\caption{Horizontal positioning performance relative to SAM segmentation.}
\label{tab:horizontal_performance}
\begin{tabular}{ccccc}
\toprule
Method &
\shortstack{Horizontal RMSE\\(m)} &
\shortstack{Improvement\\(\%)} &
\shortstack{$<3$ m\\(\%)} &
\shortstack{$<5$ m\\(\%)} \\
\midrule
Segmentation & 1.22 & ---            & 87.37 & 87.68 \\
U-Net            & 0.78 & \textbf{36.07} & 90.29 & 90.82 \\
DeepLabv3+       & 0.82 & \textbf{32.79} & 89.20 & 89.95 \\
DINOv3           & 0.81 & \textbf{33.61} & 85.99 & 86.45 \\
SAM3             & 0.86 & \textbf{29.51} & 84.65 & 85.31 \\
\bottomrule
\end{tabular}
\end{table}

Table~\ref{tab:horizontal_performance} further summarizes the horizontal positioning performance of the different methods. Compared with SAM-based segmentation, all four point-supervised models achieve lower horizontal root mean square errors (RMSE) evaluated over epochs with valid positioning solutions. U-Net achieves the lowest horizontal RMSE of 0.78 m, while the corresponding RMSEs for DeepLabv3+, DINOv3, and SAM3 are 0.82 m, 0.81 m, and 0.86 m, respectively.

To further evaluate the robustness of the proposed framework, Greedy FDE\cite{blanch2015fast} and Greedy RAIM\cite{10140135} are included as conventional fault detection methods for comparison. 
Fig.~\ref{fig:cdf} presents the cumulative distribution function (CDF) curves for the four point-supervised models, SAM-based binary segmentation, and these two conventional methods.
For horizontal error thresholds between 1 and 2 m, SAM-based segmentation and all four point-supervised models achieve substantially higher cumulative positioning success rates than Greedy FDE and Greedy RAIM, indicating that vision aided methods improve horizontal positioning accuracy. Among the evaluated methods, U-Net achieves the highest cumulative positioning success rate over the 1-3 m threshold range. Its success rates at horizontal error thresholds of 3 m and 5 m are 90.29\% and 90.82\%, respectively.

\section{Conclusion and discussion}
\label{sec:plans_conclusion}

This study developed a point-supervised visual framework for GNSS observation quality assessment in urban environments. The framework transfers measurement information from GNSS observations to a visual model through automatically generated labels at satellite projection locations.
A vision-assisted multi reference DGNSS procedure provides residual-derived fault labels and sample importance weights, enabling the model to learn a fault probability map without manual pixelwise fault annotations.
The predicted probabilities are subsequently used in the GNSS positioning solution process.

The main finding is that point supervision enables fisheye images to provide measurement quality information beyond binary sky visibility. The residual analysis shows overlapping error distributions between LOS and NLOS observations, indicating that visibility alone cannot fully describe observation reliability. This distinction is particularly relevant in vegetation-covered environments, where observations within the same non-sky category exhibit different residual magnitudes. In the examples, the point-supervised models assign different fault probabilities to these locations, with several predictions corresponding to the observed differences in measurement quality.
Observations within the same sky or non-sky category may exhibit different measurement errors.
The proposed model predicts a fault probability for each observation and uses this probability to adjust the weight of the observation in the positioning process.

Experiments with U-Net, DeepLabv3+, DINOv3, and SAM3 demonstrate the applicability of the framework to different visual architectures. Compared with SAM-based binary segmentation, the four implementations reduce horizontal positioning RMSE by 29.51-36.07\%. U-Net achieves the lowest RMSE of 0.78 m, compared with 1.22 m for the segmentation baseline. The positioning time series also show that the proposed framework can provide solutions during some intervals in which the binary segmentation approach fails. These results support the practical value of learning observation quality from measurement derived supervision and using the resulting visual information to guide GNSS observation weighting.

Future work will also explore integrating the point-supervised framework as a visual information extraction module into GNSS fault detection systems. Its predicted fault probabilities and learned visual features could be combined with GNSS observation features to exploit environmental and signal information, enabling more accurate fault detection in complex environments.

\bibliographystyle{IEEEtran}

\bibliography{refs}

\end{document}